\documentclass[twocolumn,aps,prl,showpacs]{revtex4}

\usepackage{graphicx}
\usepackage{dcolumn}
\usepackage{bm}
\usepackage[latin1]{inputenc}

\newcommand{\bra}[1]{\langle\left.{#1}\right|}
\newcommand{\ket}[1]{\left|{#1}\right.\rangle}

\begin{document}

\title{Surface states in defect-free polyatomic lattices described by a tight-binding model}
\author{Ricardo A. \surname{Pinto}}
\affiliation{Department of Electrical Engineering, University of California, Riverside, California 92521, USA}


\begin{abstract}

We report about a mechanism for surface localization, present in finite defect-free polyatomic lattices described by a tight binding model. Numerical diagonalization and degenerated perturbation theory show that there is a minimum number of atoms within each unit cell in the lattice for which surface states may exist, provided the local energy of the surface atom is different from the rest in the unit cell. It is shown that the appearance of surface states is a second-order effect in the hopping parameter. Other kinds of surface states are identified in the two-dimensional case.

\end{abstract}


\maketitle

In finite periodic lattices, the break of translational symmetry may lead to the formation of so called {\it surface states}, characterized by having wave functions which decay exponentially with the distance to the surface. This was pointed out by Tamm in his seminal work,\cite{Tamm1932PZS1} where he considered the motion of an electron in a one-dimensional semi-infinite lattice with a defect at the surface (end atom). Further studies of electronic surface states have allowed to classify them into two groups: Tamm states and the so called Shokley states. Tamm states exist in narrow-band (tight-binding) solids as a consequence of the presence of a {\it surface defect} \cite{Tamm1932PZS1,Fowler1933,Goodwin1939,Davison1968SurfaceScience11}; whereas Shokley states may exist in defect-free broad-band solids as a consequence of the crossing of energy bands\cite{Shokley1939PR56,Zak1985PRB32}. In both cases, the energy of surface states lie in the band energy gaps. There are systems where, by varying model parameters such as the surface perturbation strength and hopping, it is possible to find regions of existence (and even coexistence) of Tamm and Shokley states \cite{Foo1974PRB9,Klos2003PRB68}.
Surface states also have received much attention in the field of photonics, where the analogy between electronic transport in solids and light propagation in optical periodic media became clear \cite{Johannopoulus}. It was shown that Tamm-type surface modes may exist in the interface separating periodic and homogeneous optical media \cite{Kossel1966,Yeh1976}, where the presence of a defect at the interface determined their existence.

For many years it was thought that Tamm-type states are only possible if the lattice has a surface defect. Recently, advances in photonics have opened possibilities to excite Tamm-type states in defect-free lattices. For instance, it may be done by having a nonlinear optical medium \cite{Makris2005OptLett30,Molina2006OptLett31,nonlinOptexp}, or by periodic modulations of the lattice potential along the light propagation axis \cite{Garanovich2008PRL100}. In both ways, under certain conditions on the power and wavelength of the incoming light beam, ``effective'' defects are created at the surface, which keep light localized.
\begin{figure}[!b]
\includegraphics[width=0.95\columnwidth]{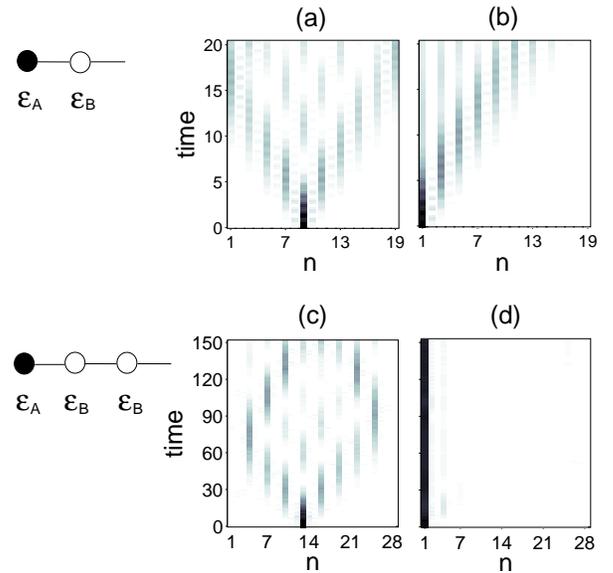}
\caption{\label{dynamics}Time evolution of the density $|\Psi_n|^2$ in (a)-(b) the $ABA$-chain, and (c)-(d), the $ABBA$-chain. Both chains have $L=10$ $A$-atoms. In each case the unit cell is shown on the left. Black and white colors correspond to maximum and minimum values of the density respectively. In (a) $N=28$, and the initial condition was $\Psi_n = \delta_{n,9}$. In (c) $N=19$, and the initial condition was $\Psi_n=\delta_{n,13}$. In (b) and (d) the initial condition was $\Psi_n=\delta_{n,1}$. In both chains all the hoppings are equal to $t$. The on-site energies are $\varepsilon_A=-5t$ and $\varepsilon_B=0$, and time is in units of $\hbar t^{-1}$.}
\end{figure} 

In this manuscript we report about a very simple way to have Tamm-type surface states in periodic defect-free lattices. One of the requirements is that the lattice must be {\it polyatomic}. It is shown that there is a lower bound for the number of atoms per unit cell (basis) for having surface states, which is three for nearest-neighbor hopping between atoms, where the local energy of the surface atom must be different from the rest in the unit cell. This condition holds even for the simplest case of a binary lattice. In this case surface states exist when two atoms of one specie are separated by, at least, two atoms of the other specie (figures \ref{dynamics}-c and d).

By using degenerate perturbation theory, it is shown that the appearance of surface states is a second-order effect in the hopping parameter. The local energies of the surface atoms receive different energy shifts in comparison to the atoms of the same specie in the bulk. Thus, they leave the corresponding energy band as hopping increases. If the basis consists of two atoms (e.g. figures \ref{dynamics}-a and b), then hybridization of local states happens at second order in the hopping parameter, and the effect is not observed. If the basis contains more than two atoms (figures \ref{dynamics}-c and d), hybridization happens at higher order. Thus, the effect is observable.
This rather simple mechanism for the existence of surface states, which was {\it hidden} during almost eighty years since Tamm's contribution, allows for practical implementations in photonic crystals, arrays of optical waveguides, and semiconductor superlattices, among probably many other possibilities.

\begin{figure}[!t]
\includegraphics[width=0.9\columnwidth]{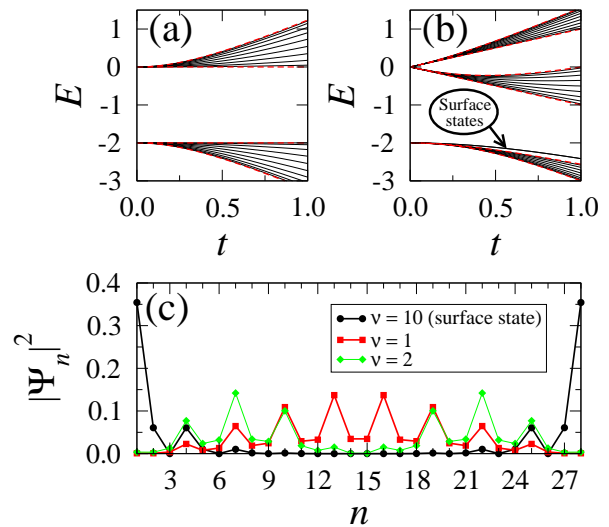}
\caption{\label{spectra}(Color online) (a) and (b), energy spectrum $E_{\nu}$ of the TB chain as a function of the hopping parameter $t$ (in arbitrary units). In (a) the chain has two atoms per unit cell ($ABA$-chain), where $N=19$ ($L=10$), $\varepsilon_A = -2$ and $\varepsilon_B = 0$. In (b) the chain has three atoms per unit cell ($ABBA$-chain), where $N=28$ ($L=10$). The red dashed lines mark the band edges of the corresponding infinite chains. (c) Spatial profile of the probability density of three eigenstates of the $ABBA$-chain belonging to the lowest-energy band [see (b)]. Here $t=1$.}
\end{figure}

We model a periodic chain of $N$ atoms with a tight binding (TB) Hamiltonian
\begin{equation}\label{eq:TBHam}
\hat{H} = \sum_{n=1}^N \epsilon_n \ket{n}\bra{n} - \sum_{n=1}^{N-1} t_{n,n+1} \left( \ket{n}\bra{n+1}
+ c.c. \right).
\end{equation}
In figures \ref{spectra}-a and b we show the energy spectra of two periodic lattices, obtained by numerical diagonalization of (\ref{eq:TBHam}) in the atomic basis $\{|n\rangle\}$. In both cases we have a binary array, where there are only two atom species $A$ and $B$ with local energies $\varepsilon_A=-2$ and $\varepsilon_B=0$ (in arbitrary units). $L$ is the number of $A$-atoms, $t_{n,n+1} = t$, and $\epsilon_1 = \varepsilon_A$.

In the case shown figure \ref{spectra}-a, the unit cell contains two atoms ($ABA$-chain, see figure \ref{dynamics} left). The spectrum consists of two energy bands. The lower and higher-energy band is formed by states having larger probability density on the odd ($A$-atoms) and even ($B$-atoms) sites respectively. The red dashed lines mark the band edges of the spectrum for the corresponding infinite chain, $E_{\pm} = [\varepsilon_A+\varepsilon_B \pm \sqrt{(\varepsilon_A-\varepsilon_B)^2+16t^2\cos^2(ka/2)}]/2$, $k$ and $a$ being the Bloch wave number and lattice constant respectively. We see that there is no energy level splitting off from any of the bands, and thus no surface state. The existence of surface states was also tested by computing the time evolution of excitations initially localized in the bulk (figure \ref{dynamics}-a) and at the surface (figure \ref{dynamics}-b) of the chain, where in both cases after a short time the wave packet spreads over the chain.

In the case shown in figure \ref{spectra}-b, the unit cell contains three atoms ($ABBA$-chain, see figure \ref{dynamics} left). The spectrum consists of three energy bands, where the lowest-energy band again is formed by states having larger probability density on the $A$-atoms, and the other two by states having larger probability density on the $B$-atoms. We may see that indeed two levels split off from the lowest energy band \cite{dispersion}. They are the surface states, characterized by an exponential decay of the probability amplitude with the distance to the surface ($A$-)atoms on the left and right ends of the chain (figure \ref{spectra}-c). An excitation initially localized in the bulk of the chain will spreads quickly (figure \ref{dynamics}-c), whereas an excitation initially localized in the surface (figure \ref{dynamics}-d) overlaps strongly with the surface states, and stays localized at the surface atom for very long times. Note also that since the surface states separate from the rest of the eigenstates of the band, they are very weakly coupled to the latter. Thus when exciting states in the bulk, the wave packet spreads over the chain, but it does not reach the surface (figure \ref{dynamics}-c), in contrast to the case where there are no surface states (figure \ref{dynamics}-a).

The existence of surface states in finite lattices may be intuitively explained as follows: The local energy at one lattice site is renormalized due to the coupling to the rest of the lattice. Since the surface atoms have different coordination number (number of atoms directly connected to them) than the atoms in the bulk, the renormalization is different for the former, effectively being impurities which lead to localization at the surface.

\begin{figure}[!t]
\includegraphics[width=2.8in]{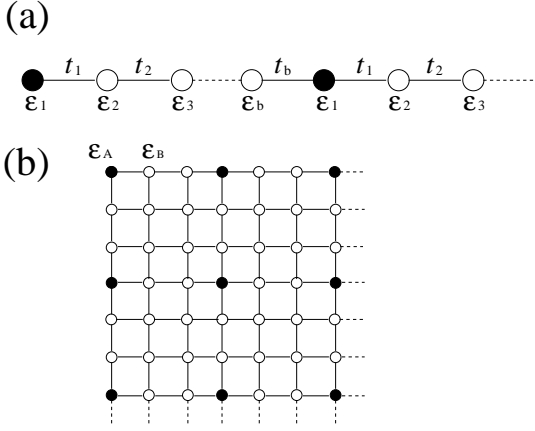}
\caption{\label{tbchain}Energy spectrum of the binary BH chain with two bosons vs eigenstate index.}
\end{figure}

{\it Analysis by degenerate perturbation theory}---
To give a description for the existence of surface states in periodic polyatomic lattices, we use degenerate perturbation theory. Let us consider a periodic chain with $N$ atoms (sites), from which $L$ of them (including the surface atoms) have local (on-site) energy $\varepsilon_1$. The primitive cell of the chain has $b>1$ atoms with on-site energies $\varepsilon_r$ ($r=1,\ldots,b$), and hopping amplitudes $t_r = t\alpha_r$, with $t_1=t$ and $\alpha_r=t_r/t_1$ (figure \ref{tbchain}-a). Thus $N=(L-1)b + 1$.

The tight binding Hamiltonian of the chain may be written as $\hat{H} = \hat{H}_0 + t\hat{V}$, where the unperturbed Hamiltonian is
\begin{eqnarray}
\hat{H}_0 &=& \sum_{r=1}^b \sum_{m=1}^M \varepsilon_r \ket{l_{m,r}}\bra{l_{m,r}} + \varepsilon_1\ket{l_{M+1,1}}\bra{l_{M+1,1}},
\end{eqnarray}
where $M=L-1$ and $l_{m,r}=b(m-1)+r$. The last term accounts for the $N$-th. site of the chain, with on-site energy $\varepsilon_1$. The hopping (perturbation) operator is
\begin{eqnarray}
\hat{V} = -\sum_{r=1}^b \sum_{m=1}^M & \alpha_r & \Big( \ket{l_{m,r}}\bra{l_{m,r}+1}
+ \;c.c. \Big).
\end{eqnarray}

At $t=0$, the eigenstates of the system are the localized basis states $\{|{l_{m,r}}\rangle\}$. We are interested in the continuation of the states $\ket{l_{m,1}}$ for $t>0$, since they include the surface basis states $\ket{1}$ and $\ket{N}$. We assume that $\varepsilon_{r\neq 1} \neq \varepsilon_1$. Therefore, the states of interest are $(M+1)=L$-fold degenerated. The zeroth-order wavefunction is written as:
\begin{equation}
|\Psi_{l_{m,1}}^{(0)}\rangle = \sum_{m' = 1}^{M+1} C_{m',m} \ket{l_{m' ,1}}.
\end{equation}

The first non-zero correction to the eigenenergy comes in second order in the hopping parameter $t$, and it is
obtained from the equation
\begin{equation}\label{eq:secondorderenergy}
\bra{l_{m' ,1}}\hat{V}|\Psi_{l_{m,1}}^{(1)}\rangle = E_{l_{m,1}}^{(2)} C_{m',m},
\end{equation}
where $m,m'=1,\ldots,M+1$, and $|\Psi_{l_{m,1}}^{(1)}\rangle$ is the first-order correction to the wavefunction. The left hand side of (\ref{eq:secondorderenergy}) is equal to
\begin{eqnarray}\label{eq:lefthand}
& &  - \sum_{m''=2}^{M+1}\frac{\alpha_b^2}{\varepsilon_b - \varepsilon_1} C_{m'',m} \delta_{m', m''} \nonumber \\
& & - \sum_{m''=1}^{M}\frac{\alpha_1^2}{\varepsilon_2 - \varepsilon_1} C_{m'',m} \delta_{m', m''} \nonumber \\
& & - \sum_{m''=2}^{M+1}\frac{\alpha_b\alpha_{b-1}}{\varepsilon_b - \varepsilon_1} C_{m'',m} \delta_{m'+2/b, m''} \nonumber \\
& & - \sum_{m''=1}^{M}\frac{\alpha_1\alpha_{2}}{\varepsilon_2 - \varepsilon_1} C_{m'',m} \delta_{m'-2/b, m''}.
\end{eqnarray}
Thus, Eq. (\ref{eq:secondorderenergy}) is equivalent to the eigenvalue equation 
$
\mathcal{H}\mathbf{C}_{m} = E_{l_{m,1}}^{(2)}\mathbf{C}_{m}
$,
where $\mathbf{C}_{m} = (C_{1,m}\; C_{2,m}\;\ldots \; C_{M+1,m})^t$. The diagonal elements of the $(M+1)\times (M+1)$ matrix $\mathcal{H}$ are
\begin{equation}\label{eq:diag}
\mathcal{H}_{m',m'} = \left\{ \begin{array}{ll}
-\frac{\alpha_1^2}{\varepsilon_2-\varepsilon_1} & \textrm{if $m'=1$},\\
\\
-\left(\frac{\alpha_b^2}{\varepsilon_b-\varepsilon_1} + \frac{\alpha_1^2}{\varepsilon_2-\varepsilon_1}\right) & \textrm{if $2\leq m'\leq M$},\\
\\
-\frac{\alpha_b^2}{\varepsilon_b-\varepsilon_1} & \textrm{if $m'=M+1$.}
\end{array} \right.
\end{equation}
The only off-diagonal elements that may have non-zero values are
\begin{eqnarray}\label{eq:offdiag}
\mathcal{H}_{m',m'-2/b} &=& \frac{\alpha_1\alpha_2}{\varepsilon_2 - \varepsilon_1} \nonumber \\
\nonumber \\
\mathcal{H}_{m',m'+2/b} &=& \frac{\alpha_b\alpha_{b-1}}{\varepsilon_b - \varepsilon_1}.
\end{eqnarray}

From Eqs. (\ref{eq:diag}) and (\ref{eq:offdiag}) we see that if $b>2$, the off-diagonal matrix elements vanish, and the second-order correction to the energy is given by (\ref{eq:diag}). Thus, at second order in the hopping parameter, the degeneracy is not completely lifted. The two eigenvalues $E_{1}^{(2)}=\mathcal{H}_{1,1}$ and $E_{N}^{(2)}=\mathcal{H}_{M+1,M+1}$ correspond to the surface states, which split off from the remaining degenerated levels which are not surface states. Up to second order, the eigenenergies are, for $m=1$ and $m=M+1=L$,
\begin{eqnarray}\label{eq:ensurf}
E_1 & \simeq & \varepsilon_1 - \frac{\alpha_1^2}{\varepsilon_2-\varepsilon_1} t^2, \\
\nonumber \\
E_N & \simeq & \varepsilon_1 - \frac{\alpha_b^2}{\varepsilon_b-\varepsilon_1} t^2.
\end{eqnarray}
For $m\neq 1,M+1$,
\begin{equation}\label{eq:ennonsurf}
E_{b(m-1)+1} \simeq  \varepsilon_1 - \left(\frac{\alpha_b^2}{\varepsilon_b-\varepsilon_1} + \frac{\alpha_1^2}{\varepsilon_2-\varepsilon_1}\right) t^2,
\end{equation}
The degeneracy of the remaining non-surface levels (\ref{eq:ennonsurf}) will be lifted at higher order $\mathcal{O}(t^b)$, and will form a band of bulk states.
If $b=2$, then there are non-zero off-diagonal matrix elements $\mathcal{H}_{m',m'\pm 1}$. Thus, at second order the degeneracy is lifted and we have a band of energy levels, and no surface states split off from the band.
Finally, we see from (\ref{eq:ensurf}) that the existence of surface states is subject to the condition $\varepsilon_{r\neq 1} \neq \varepsilon_1$, no matter whether the on-site energies $\varepsilon_{r\neq 1}$ are equal or not; and that the surface states are not degenerated if $\varepsilon_b \neq \varepsilon_2$ or $\alpha_b \neq \alpha_1$. 
\begin{figure}[!t]
\includegraphics[width=1.8in]{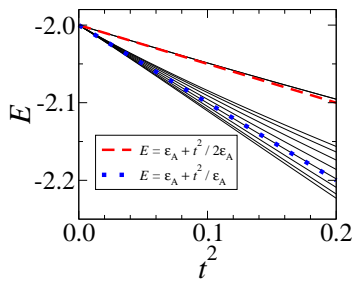}
\caption{\label{ptspectrum}(Color online) Energy spectrum vs the squared hopping parameter around the lowest-energy band (in arbitrary units), obtained by numerical diagonalization for the $ABBA$-chain ($b=3$) with $N=28$, $\varepsilon_A=-2$, and $\varepsilon_B = 0$ (see figure \ref{spectra}-b). The surface states split off from the rest. The thick dashed and dotted lines are the results from Eqs. (\ref{eq:ensurf}--\ref{eq:ennonsurf}) with $\varepsilon_1\equiv\varepsilon_A$, $\varepsilon_{2,b}\equiv\varepsilon_B$, and $\alpha_r=1$ ($r=1,2,3$).}
\end{figure} 

In figure \ref{ptspectrum} we show the energy spectrum of the $ABBA$-chain around the lowest-energy band, obtained by numerical diagonalization of the Hamiltonian (\ref{eq:TBHam}) with $\varepsilon_A = -2$ and $\varepsilon_B=0$ (thin solid lines), where we see the splitting off of the surface states from the upper band-edge. Up to second order in the hopping parameter, degenerated perturbation theory (thick dashed and dotted lines) nicely describes the appearance of the surface states. These results are similar to those obtained by Pinto {\it et al} for a many-particle model, where by using perturbative arguments the system was reduced to an effective model describing one particle in a polyatomic lattice with a particular structure in the primitive cell \cite{Pinto2009}.

{\it Surface states in two-dimensions}--- Having demonstrated the existence of surface states in finite one-dimensional lattices, the extension to the two-dimensional case is straightforward, where a similar analysis using degenerate perturbation theory may be carried out. For the sake of simplicity, we again considered a binary array (only two on-site energies) and equal hopping parameters (figure \ref{tbchain}-b), where the impurity atoms with on-site energy $\varepsilon_A$ are separated by $b$ atoms with on-site energy $\varepsilon_B$ along the horizontal and vertical directions. The result is that again surface states may exist if $b>2$, and that there are two groups of surface states: One group are {\it corner states}, which are localized at the corners of the lattice. Up to second order in the hopping parameter, they are four-fold degenerated, with eigenenergy
$
E_{corner} \simeq \varepsilon_A - 2t^2/(\varepsilon_B-\varepsilon_A).
$
The other group are {\it edge states}, which are localized along the edges (excluding the corners) of the lattice. They are also four-fold degenerated, with eigenvalue
$
E_{edge} \simeq \varepsilon_A - 3t^2/(\varepsilon_B - \varepsilon_A).
$
The other eigenstates are $(L-2)^2$-fold degenerated bulk states with eigenenergy $E_{bulk} \simeq \varepsilon_A - 4t^2/(\varepsilon_B-\varepsilon_A)$, which will hybridize at order $\mathcal{O}(t^b)$.

The above-described two groups of surface states obtained by perturbation theory are consistent with expectation from renormalization arguments, since the coordination number of the atoms along the edges is different from the atoms in the bulk; and in turn the coordination number of the corner atoms is different from the  one for atoms along the edges.

In Summary, we reported the existence of single-particle surface states in finite defect-free polyatomic lattices. We have shown that there is a minimum number of atoms per unit cell (basis) for which such surface states may exist, which is three in the one-dimensional case. This number gives the minimum basis along each direction in the two-dimensional case, where two kinds of surface states were identified. We expect similar results in the three-dimensional case. The lower bound in the basis is because the appearance of surface states is a second-order effect in the hopping parameter, a fact which had remained hidden for long time since Tamm's contribution. The rather simple conditions for the existence of surface states described here allows for practical implementations. Although we presented results for the electronic case, we expect that they hold in the optical case as well, where surface localization is nowadays object of intensive research.

\acknowledgments
We acknowledge S. Flach for useful discussions, and acknowledge the hospitality of the Max Planck Institute for the Physics of Complex Systems in Dresden, Germany, where most of this work was done.


\begin{thebibliography}{99}

\bibitem{Tamm1932PZS1}
I. Tamm, Phys. Z. Sowjetunion {\bf 1}, 733 (1932).

\bibitem{Fowler1933}
R. H. Fowler, Proc. R. Soc. Londno, Serv. A {\bf 141}, 56 (1933); S. Rijanow, Z. Phys. {\bf 89}, 806 (1934).

\bibitem{Goodwin1939}
E. T. Goodwin, Proc. Camb. Phil. Soc. {\bf 35}, 205 (1939); {\bf 35}, 221 (1939); {\bf 35}, 232 (1939).

\bibitem{Davison1968SurfaceScience11}
S. G. Davison and J. Grindlay, Surf. Sci. {\bf 11}, 99 (1968); S. G. Davison, T. Y. Ling, and U. S. Ghosh, J. Phys. Chem. Solids {\bf 28}, 1921 (1967).

\bibitem{Shokley1939PR56}
W. Shokley, Phys. Rev. {\bf 56}, 317 (1939).

\bibitem{Zak1985PRB32} J. Zak, Phys. Rev. B {\bf 32}, 2218 (1985).

\bibitem{Foo1974PRB9}
E-Ni Foo, How-Sen Wong, Phys. Rev. B {\bf 9}, 1857 (1974).

\bibitem{Klos2003PRB68}
J. Klos and H. Puszkarski, Phys. Rev. B {\bf 68}, 045316 (2003). 

\bibitem{Johannopoulus}
J. D. Johannopoulus, R. D. Meade, J. N. Winn, {\it Photonic crystals, molding the flow of light}, (Lawrenceville, NJ, 1995).

\bibitem{Kossel1966}
D. Kossel, J. Opt. Soc. Am. {\bf 56}, 1434 (1966); J. A. Arnaud and A. A. M. Saleh, Appl. Opt. {\bf 13}, 2343 (1974).

\bibitem{Yeh1976}
P. Yeh, A. Yariv, and Chi-Shain Hong, J. Opt. Soc. Am. {\bf 67}, 424 (1977). P. Yeh, A. Yariv, A. Y. Cho, Appl. Phys. Lett. {\bf 32}, 104 (1978).

\bibitem{Makris2005OptLett30}
K. G. Makris, S. Suntsov, D. N. Christodoulides, G. I. Stegeman, and A. Hach\'e, Opt. Lett. {\bf 30}, 2466 (2005).

\bibitem{Molina2006OptLett31}
M. I. Molina, R. A. Vicencio, Yu. S. Kivshar, Opt. Lett. {\bf 31}, 1693 (2006).

\bibitem{nonlinOptexp}
S. Suntsov, K. G. Makris, D. N. Christodoulides, G. I. Stegeman, A. Hach\'e, R. Morandotti, H. Yang, G. Salamo, M. Sorel, Phys. Rev. Lett. {\bf 96}, 063901 (2006); E. Smirnov, M. Stepi\'c, C. E. R\"uter, D. Kip, V. Shandarov, Opt. Lett. {\bf 31}, 2338 (2006); C. R. Rosberg, D. N. Neshev, W. Krolikowski, A. Mitchell, R. A. Vicencio, M. I. Molina, and Yu. S. Kivshar, Phys. Rev. Lett. {\bf 97}, 083901 (2006).

\bibitem{Garanovich2008PRL100}
I. L. Garanovich, A. A. Sukhorukov, and Yu. S. Kivshar, Phys. Rev. Lett. {\bf 100}, 203904 (2008).

\bibitem{dispersion}
The band edges for the $ABBA$-chain in figure \ref{spectra}-b were obtained by numerically finding the roots of the cubic equation which gives the dispersion relation $E(k)$, for $k=0$ and $ka=\pi$, at each value of the hopping parameter $t$.

\bibitem{Pinto2009}
Ricardo A. Pinto, Masudul Haque, and Sergej Flach, Phys. Rev. A {\bf 79}, 052118 (2009).

\end{thebibliography}
\end{document}